\documentclass[preprint]{aastex}
\usepackage{amsmath}
\usepackage{hyperref}

\usepackage[english]{babel}
\usepackage{graphicx}
\usepackage{latexsym}
\usepackage{amsmath,amssymb}
\usepackage{natbib}
\usepackage{dcolumn}
\usepackage{bm}
\usepackage{calrsfs}
\usepackage{ulem}
\usepackage[usenames]{color}
\usepackage{xspace}

\newcommand{\nn}{\mbox{} \nonumber \\ \mbox{} }
\newcommand{\ba}{\begin{eqnarray}}
\newcommand{\ea}{\end{eqnarray}}
\newcommand{\om}{\omega}

\newcommand{\Bf}{{magnetic field}}

\newcommand{\NS}{neutron star}

\newcommand{\Lf}{{Lorentz factor}}

\newcommand\eg{\textit{e.g.}}

\def\be{\begin{equation}}
\def\ee{\end{equation}}

\begin{document}

\title{Wind-powered   afterglows of gamma-ray bursts: flares, plateaus and steep decays}


\author{Yonggang Luo, Maxim Lyutikov\\
Department of Physics and Astronomy, Purdue University, \\
West Lafayette, IN
47907-2036, USA}

\begin{abstract}
Afterglows of gamma-ray bursts  often show flares, plateaus, and sudden intensity drops:  these temporal features are difficult to explain as coming from the  forward shock. We calculate radiative properties of  early GRB afterglows with the dominant contribution from the reverse shock (RS) propagating in an ultra-relativistic (pulsar-like)  wind produced by the  long-lasting central engine. RS emission occurs in the fast cooling regime - this  ensures high radiative efficiency  and allows fast intensity variations.  We demonstrate that: (i) mild wind power, of the order of  $\sim 10^{46}$ erg s$^{-1}$, can reproduce the  afterglows' plateau phase; 
(ii)   termination of the  wind 
can  produce sudden steep decays;
 (iii) mild variations in the wind luminosity can produce short-duration afterglow flares.
 \end{abstract}
\maketitle 

\section{Introduction}

Gamma-ray bursts (GRBs) are produced in relativistic explosions \citep{1986ApJ...308L..43P,2004RvMP...76.1143P} that 
generate two shocks: forward shock and reversed shock. The standard fireball model \citep{1992MNRAS.258P..41R,1995ApJ...455L.143S,1999PhR...314..575P,2006RPPh...69.2259M} postulates that 
the prompt emission is produced by internal  dissipative  processes  within the flow: collisions of matter-dominated shells, \cite{1999PhR...314..575P}, or reconnection events \citep{2006NJPh....8..119L}). The   afterglows, according to the fireball model, are generated in 
the external relativistic blast wave.

One of the most surprising results of the Swift observations of the early  afterglow is the presence of temporal structures not expected in the standard model: 
plateaus and flares  \citep{2006ApJ...642..389N}, and 
 sudden steep decays, \eg\ in  GRB 070110 \citep{2007ApJ...665..599T}. These features are hardly consistent with the standard fireball model, as discussed by 
 \cite{Lyutikov:2009,2017ApJ...835..206L}. 
 
 The origin of sudden drops in afterglow light curves  is especially  mysterious. As an example,
  GRB 070110 starts with a normal prompt emission, followed 
by an early decay phase until approximately  100 seconds,  and a plateau until $\sim 10^4$ s.
 At  about $2 \times 10^4$ seconds, the light 
curve of the afterglow of GRB 070110 drops suddenly with a temporal  slope $> 7$ \citep{2007GCN..6008....1S,2007GCN..6014....1K,2007GCNR...26....2K,2007ApJ...665..599T}.

Such an abrupt steep decay  in afterglow light curves is   inconsistent with the standard fireball model. Such sharp drops require (at the  least) that  the  emission from the forward shock (FS) switches off instantaneously.  This is  impossible.  First,  the microphysics of shock acceleration is not expected to change rapidly (at least we have no arguments why it should). The variations of  hydrodynamic properties of the FS, as they translate to radiation,  are also expected to produce smooth variations.   For example, as a model problem consider a relativistic  shock that breaks out from a denser medium (density $n_1$) into the less dense one (density $n_2\ll  n_1$). In the standard fireball model total  synchrotron power $P_s$ per unit area of the shock   scale as \citep{2004RvMP...76.1143P}
\ba &&
P_s \propto n \Gamma^2 \gamma^{\prime 2} B^{\prime 2} \propto n^2 \Gamma^6
\nn &&
\gamma^{\prime} \propto \Gamma
\nn && 
B^{\prime}  \propto \Gamma \sqrt{n}
\ea
where $\Gamma$ is the \Lf\ of the shock,  $\gamma^{\prime} $ is the \Lf\ of accelerated particles.

 Importantly, if  a shock breaks out from a dense medium into the rarefied one, with $n_2\ll  n_1$, it {\it accelerates} to approximately $\Gamma_2 \approx \Gamma_1^2$, as the post-shock internal energy in the first medium is converted into bulk motion  \citep {1971PhRvD...3..858J,2010PhRvE..82e6305L}. 
Thus a change in power and peak frequency scale as
\be 
\frac{P_{s,2}}{P_{s,1}} ={\Gamma_1^6} \left(  \frac{n_2}{n_1}\right)^2
\ee
Thus, even though we assumed $n_2 \ll n_1$, the synchrotron emissivity in the less dense medium is largely  compensated by the increase of the \Lf. 
 Since the expected \Lf\ at the  time of sharp drops is $ \Gamma_1 \sim$ few tens, suppression of emission from the forward shock requires the unrealistically large decrease of  density. 

As we discuss in this paper, the abrupt declines in afterglow curves can be explained if emission originates in the ultra-relativistic  reverse shock of a long-lasting engine.
\cite{2017ApJ...835..206L} \citep[see also][]{2017PhFl...29d7101L,2020arXiv200413600B} developed a model of early GRB afterglows with dominant X-ray contribution from the highly magnetized ultra-relativistic  reverse shock (RS), an analog of the pulsar wind termination shock.
The critical point is that emission from the RS in highly magnetized pulsar-like wind  occurs in the fast cooling regime. Thus it  reflects {\it instantaneous} wind power, not accumulated mass/energy, as in the case of the forward shock. Thus, it is more natural to produce fast variation in the  highly magnetized RS.

The model by \cite{2017ApJ...835..206L} has several key features. (i) the high energy X-ray and the optical synchrotron emission from the RS particles occur in the fast cooling regime - this ensures efficient conversion of the wind power into radiation and thus can  account for rapid  variability  due to changes in the wind properties.; (ii)  plateaus -- parts of afterglow light curves that show slowly decreasing  power -- are a natural consequence of the RS emission. We study these effects in more detail in the present paper.

In this  work, we explore a  model  that most of the early  X-ray afterglow emission comes from 
the RS of a long-living central engine. This allows us to resolve the problems of plateaus, sudden intensity drops, and flares. Qualitatively, first,  at early times, a large fraction of the wind power is radiated: this explains the plateaus. Second, if the wind terminates, so that the  emission from RS  ceases  instantaneously,
 this will lead to a sharp decrease in observed flux (since particles are cooling fast). Third, variations of the wind intensity can  produce observed flares.


\section{Emission from relativistic termination shock}
\label{rad}

\subsection{Wind dynamics}
 
Following \cite{2017ApJ...835..206L}, we assume that  a  powerful pulsar is born in the initial GRB explosion. The pulsar produces a highly magnetized and highly relativistic  pulsar-like wind that shocks against the expanding ejecta. Thus,  the system constitutes a  relativistic double explosion \citep{2017PhFl...29d7101L,2020arXiv200413600B}. 

Let the central source produce luminosity per solid angle  $dL/d\Omega$ that is carried by particles and \Bf, 
\ba &&
\frac{dL}{d\Omega} =  (\rho '  + \frac{B^{\prime, 2}}{4\pi} ) r^2 \gamma_{w}^2= (1+\sigma) \rho 'r^2 \gamma_{w}^2
\nn &&
\sigma =  \frac{B^{\prime, 2}}{4\pi \rho ' } 
\ea
where $\rho '$ is plasma density, $B'$ is the toroidal magnetic field, and $\gamma_{w}$ is the Lorentz factor of the wind; the speed of light was set to unity. In this work, we denote 
primed  variables  in the fluid frame. 

\begin{figure}[h!]
  \centering
  \includegraphics[width=0.8\textwidth]{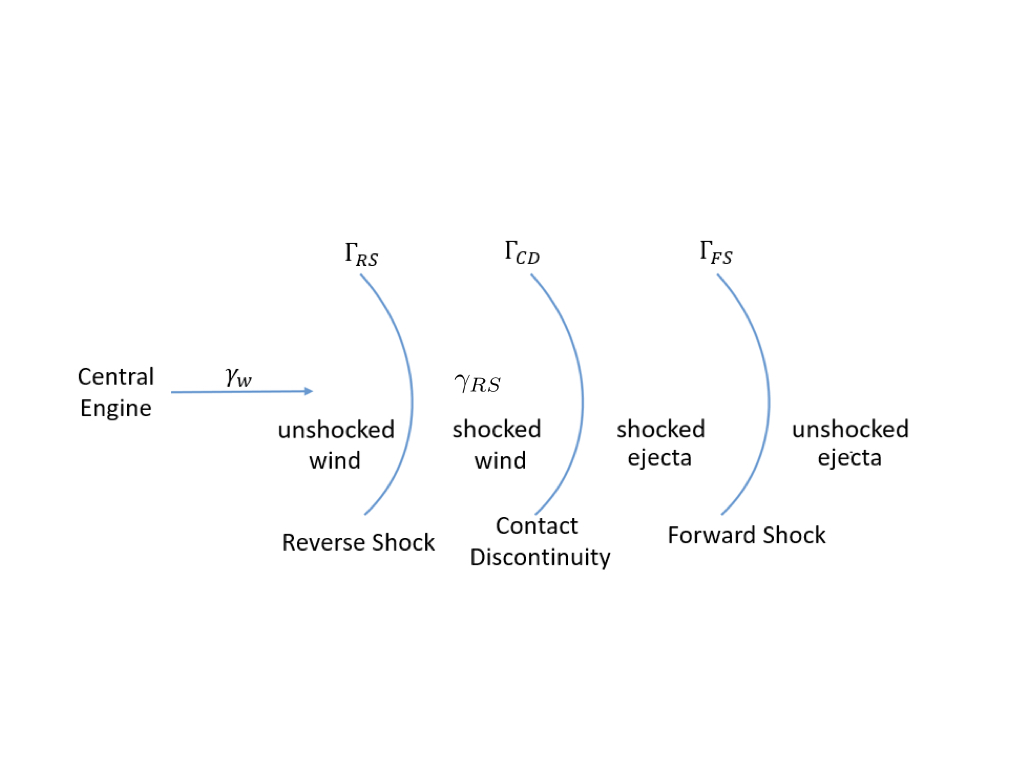}
  \caption{Structure of the   wind-ejecta flow. The  wind moving with \Lf\ $\gamma_w$ terminates at the reverse shock (RS), which is  moving with $\Gamma_{RS}$.
  The post-RS flow is moving  with $\gamma_{RS}\approx \Gamma_{RS} \approx \Gamma_{CD}$ (the \Lf\ of the contact discontinuity separating wind and ejecta material). }
  \label{rsfs}
 \end{figure}

In a pulsar paradigm, the wind is highly magnetized, $\sigma  \gg 1$, and extremely relativistic, $ \gamma_w \sim 10^4-10^6$ \citep{kennel_coroniti_84,langdon_88,1992ApJ...390..454H}.
This highly magnetized wind  shocks against relativistically expanding ejecta, Fig. \ref{rsfs}.  The emission is produced in the shocked wind moving with the \Lf\ $\gamma _{RS} \approx \Gamma_{RS} \approx \Gamma_{CD}$,  where $\gamma _{RS}$  is the \Lf\ of the post-reverse shock flow, $\Gamma_{RS}$ is the \Lf\ of the reverse shock (RS) and $\Gamma_{CD}$ is the \Lf\ of the contact discontinuity between the wind and the preceding ejecta.

The dynamics of the double relativistic  explosions are somewhat complicated \citep{2017PhFl...29d7101L,2020arXiv200413600B}. The second shock sweeps-up the tail material from the initial explosion. Thus, the dynamics of the second shock depends on the internal  structure of the post-first shock flow, and  the wind power; all pressure relations are highly complicated by the relativistic and time-of-flight  effects. 
Under certain conditions, the flow is approximately self-similar.

To avoid the mathematical complications, and to demonstrate the essential physical effects most clearly, we assume a simplified dynamics of the second shock, allowing it to propagate with constant velocity. Thus, in the frame of the shock, the \Bf\ decreases linearly with time,
\be
B^{\prime}=B_0^{\prime} \frac{t_0^{\prime}}{t^{\prime}}
\label{B}
\ee
where time $t_0^{\prime}$ and  magnetic field is $B_0^{\prime}$ are some constants.
In the following, we  assume that the RS starts to accelerate particles at time $t_0^{\prime}$, and  we calculate the emission properties of particles injected at the wind termination shock taking into account    radiative and adiabatic losses.

\subsection{Evolution of  the distribution function}

As the  wind generated by the long-lasting engine  starts to interact with the tail part of the flow generated by the initial explosion, the RS forms in the wind, see Fig. \ref{rsfs}. Let's assume that the RS accelerates particles with  a power-law distribution, 
\begin{eqnarray}
f\left(\gamma^{\prime},t_i^{\prime}\right) \propto   {\gamma^{\prime}}^{-p} \Theta(\gamma^{\prime}- \gamma_{\text{min}}^{\prime})
\end{eqnarray}
where $t_i'$ is the injection time, $\Theta$ is the step-function, $\gamma$ is the Lorentz factor of the particles, and $\gamma_{\text{min}}^{\prime}$ is the minimum Lorentz factor of the injected particles. $\gamma_{\text{min}}^{\prime}$ can be estimated as \citep{1984ApJ...283..710K}
\be
\gamma_{\text{min}}^{\prime} \sim \gamma_{RS}  \sim  \gamma_{w}/2\Gamma_{RS}
\label{min}
\ee
(We stress that in the pulsar-wind paradigm the minimal \Lf\ of accelerated particles  $\gamma_{\text{min}}^{\prime}$ scales differently from the matter-dominated  fireball case.)

The accelerated particles produce synchrotron emission in the ever-decreasing \Bf, while also experiencing adiabatic losses. Synchrotron losses are given by the standard relations \citep[\eg][]{1999acfp.book.....L}. To take account of adiabatic losses we note that the conservation of the first adiabatic invariant (constant magnetic flux through the cyclotron orbit) gives
\be
\partial_ {t^{\prime}} \ln \gamma^{\prime} = \frac{1}{2} \partial_ {t^{\prime}} \ln B^{\prime}
\ee
(thus, we assume that that \Bf\ is dominated by the large-scale toroidal field).

Using Eqn. (\ref{B}) for the evolution of the field,  the evolution of a particles' \Lf\ follows
\begin{eqnarray}  
&&
\frac{{d}\gamma^{\prime}}{{dt^{\prime}}}=-\frac{\tilde{C}_1 {B_0^{\prime}}^2 {\gamma^{\prime}}^2}{{t^{\prime}}^2}-\frac{\gamma^{\prime}}{2 t^{\prime}}
\nn && 
\tilde{C}_1=\frac{\sigma_T {t_0^{\prime}}^2}{6 {\pi  m _e c}}
\label{gammapde}
\end{eqnarray}
where $\sigma_T$ is the Thomson cross-section and $t_0^{\prime}$ is some reference  time.

Solving for the evolution of the particles' energy in the flow frame,
\begin{eqnarray}
\frac{1}{\gamma^{\prime}}=\frac{2 \tilde{C}_1 {B_0^{\prime}}^2}{3 t^{\prime}} \left(\left(\frac{t^{\prime}}{{t}_i^{\prime}}\right)^{3/2}-1\right)+\frac{1}{\gamma_i^{\prime}}\sqrt{\frac{t^{\prime}}{t_i^{\prime}}},
\end{eqnarray}
we can derive the evolution of a distribution function (the Green's function) \citep[\eg][]{1962SvA.....6..317K,1984ApJ...283..710K}
\begin{eqnarray} &&
G(\gamma^{\prime},t^{\prime},t_i^{\prime})=
\left\{ 
\begin{array}{cc}
{\gamma^{\prime}}^{-p} \left(\frac{t_i^{\prime}}{t^{\prime}}\right)^{\frac{p-1}{2}} \left(1-\frac{2}{3} \tilde{C}_1 {B_0^{\prime}}^2 \gamma_w^{\prime} \sqrt{t^{\prime}} \left(\frac{1}{{t_i^{\prime}}^{3/2}}-\frac{1}{{t^{\prime}}^{3/2}}\right)\right)^{p-2}, & {\gamma_{\text{low}}^{\prime}<\gamma^{\prime}<\gamma_{\text{up}}^{\prime}} \\
 0, & {else} \\
\end{array}
\right.
\nn &&
\frac{1}{\gamma_{\text{low}}^{\prime}} = \frac{2 \tilde{C}_1 {B_0^{\prime}}^2}{3 t^{\prime}} \left(\left(\frac{t^{\prime}}{t_i^{\prime}}\right)^{3/2}-1\right)+\frac{1}{\gamma_{min}^{\prime}}\sqrt{\frac{t^{\prime}}{t_i^{\prime}}}
\nn && 
\frac{1}{\gamma_{\text{up}}^{\prime}} = \frac{2 \tilde{C}_1 {B_0^{\prime}}^2}{3 t^{\prime}} \left(\left(\frac{t^{\prime}}{t_i^{\prime}}\right)^{3/2}-1\right)
\end{eqnarray}
where $\gamma_{\text{low}}'$ is a lower bound of Lorentz factor due to minimum Lorentz factor at injection and $\gamma_{\text{up}}'$ is an upper bound of Lorentz factor due to cooling.

Once we know the evolution of the distribution function injected at time $t_i'$,  we can use the Green's function to derive the total distribution function  by integrating over the injection times 
\be
{N}(\gamma^{\prime},t^{\prime}) \propto  \int _{t_i'}^{t'}  \dot{n} (t_i') G(\gamma^{\prime},t^{\prime},t_i^{\prime})dt_i^{\prime}
\ee
where $ \dot{n} (t_i') $ is  the injection rate (assumed to the constant below).

\subsection{Observed intensity}

The intensity  observed at each moment depends on the intrinsic luminosity, the geometry of the flow, relativistic,  and time-of-flight effects \citep[\eg][]{1996ApJ...473..998F,2003NewA....8..495N,2004RvMP...76.1143P}. 


The intrinsic emissivity at time $t^\prime$ depends on the  distribution function $N$ and synchrotron power $P_\omega$:
\be
L^{\prime}(\omega^{\prime}, t^{\prime})=\int \int {N_A(\gamma^{\prime}, t^{\prime}) P_\omega(\omega^{\prime})} \, d\gamma^{\prime} dA^{\prime}
\ee
where $N_A$, the number of particles per unit area, is defined as $N_A = N / A = N / (2 \pi {r^{\prime}}^2 (1- \cos \theta_j))$,
 $P(\omega^{\prime})$ is the power per unit frequency emitted by each electron, and $dA^{\prime}$ is the surface differential \citep[unlike][we do not have extra $\cos \theta$ in the expression for the area since we use volumetric emissivity, not emissivity from a surface]{1996ApJ...473..998F}.

We assume that the observer is located on the symmetry axis and that   the active part of the RS occupies angle $\theta_j$ to the line of sight.
The emitted power is then
\be
L^{\prime}(\omega^{\prime}, t^{\prime})= \int_{0}^{\theta_j} \int _{\gamma_{\min }^{\prime}}^{\infty }  N_A(\gamma^{\prime} ,t^{\prime}) P(\omega^{\prime}) d\gamma^{\prime}  2 \pi  {r^{\prime}}^{2} \sin(\theta) d\theta
\label{luminosity}
\ee

Photons  seen  by a distant observer at times $T_{ob}$ are emitted at  different radii and  angles $\theta$. To take account of the time of flight effects, we note that the distance between the initial explosion point and an emission point $(r^{\prime}, \theta)$ is $r^{\prime}= v t^{\prime}=v T_{ob} (1-\beta \cos(\theta))^{-1} \gamma_{RS}^{-1}$, where $T_{ob}$ is the observed time. Supposed that a photon was emitted from the distance $r^{\prime}$ and angle $\theta = 0$ at time $t^{\prime}$, and at the same time, the other photon was emitted from the distance $r^{\prime}$ and any arbitrary angle $\theta = \theta_i < \theta_j$. These two photons will be observed at time $T_0$ and $T_{\theta_i}$, then the relation between $T_0$ and $T_{\theta_i}$ is given by: 
\begin{eqnarray}
r^{\prime}=v t^{\prime}=\frac{v T_0}{(1-\beta)\gamma_{RS}}=\frac{v T_{\theta_i}}{(1-\beta \cos(\theta_i))\gamma_{RS}}
\label{surface}
\end{eqnarray}
where, the time $t^{\prime}$ measured in the fluid frame, and the corresponding observe time $T_{ob}$, is a function of $\theta$ and $t^{\prime}$: 
\begin{eqnarray}
T_{ob} = t \left(1 - \beta \cos \theta\right)= t^{\prime} \left(1 - \beta \cos \theta\right) \gamma_{RS}
\label{ob}
\end{eqnarray}

Taking the derivative of Eqn. (\ref{ob})  we find 
\begin{eqnarray}
\sin(\theta) d\theta = - \frac{T_{\text{ob}}}{{t^{\prime}}^2 \beta \gamma_{RS}} dt^{\prime} \approx - \frac{T_{\text{ob}}}{{t^{\prime}}^2 \gamma_{RS}} dt^{\prime}
\label{theta}
\end{eqnarray}
Substitute the relation (\ref{theta}) into  (\ref{luminosity}), the observed  luminosity becomes
\begin{eqnarray}
L^{\prime}(T_{ob},\omega^{\prime}) \approx \int_{t_{\theta^{\prime}=0}^{\prime}}^{t_{\theta^{\prime}=\theta_j}^{\prime}} \int _{\gamma_{\min }^{\prime}}^{\infty } \frac{ 2 \pi c^2 T_{ob}}{\gamma_{RS}} \times N_A(\gamma^{\prime} ,t^{\prime}) P(\omega^{\prime}) d\gamma^{\prime} dt^{\prime}
\label{drop}
\end{eqnarray}

To understand the Eqn. (\ref{drop}), the radiation observed at $T_{ob}$ corresponds to the emission angle from $0$ to $\theta_{j}$, which also corresponds to the emission time $t_{\theta^{\prime}=0}^{\prime} ={T_{ob}}/{(1-\beta)\gamma_{RS}}$ to $t_{\theta^{\prime}=\theta_j}^{\prime} ={T_{ob}}/{(1-\beta \cos \theta_j)\gamma_{RS}}$. So we need to integrate the emissivity function over the range of the emission angle, or integrate the emissivity function over the range of the emission time from $t_{\theta^{\prime}=0}^{\prime}= {T_{ob}}/{(1-\beta)\gamma_{RS}}$ to $t_{\theta^{\prime}=\theta_j}^{\prime}={T_{ob}}/{(1-\beta \cos \theta_j)\gamma_{RS}}$.

Finally, taking into account Doppler effects  (Doppler shift $\omega = \delta \omega^{\prime}$ and the intensity boost $I_{\omega} \left(\omega\right) = \delta^3 I_{\omega^{\prime}}^{\prime} \left(\omega^{\prime}\right)$; where  $\delta$ is the Doppler factor $\delta = 1/({\gamma_{RS} \left(1 - \beta \cos \theta\right)}) $), 
substitute the relation $t^{\prime}={T_{\text{ob}}}/{(1-\beta \cos(\theta))\gamma_{RS}}$ into Eqn.(\ref{drop}) we finally arrive at the equation for the observed  spectral luminosity:
\begin{eqnarray}
F_\om  = \int_{\frac{T_{\text{ob}}}{(1-\beta \cos(\theta_j))\gamma_{RS}}}^{\frac{T_{\text{ob}}}{(1-\beta)\gamma_{RS}}} \int _{\gamma_{\min }^{\prime}}^{\infty } \frac{1}{2 \gamma_{RS}}  c^2  D^{-2} T_{\text{ob}} \delta^3 N_A P(\omega/\delta) d\gamma^{\prime} dt^{\prime}
\label{final}
\end{eqnarray}
where $D$ is the distance to the GRB.

\section{Results}
\label{result}

In the following, we apply the general relations  derived above to the  three specific problem: (i) origin of  plateaus in afterglow light curves and (ii) sudden drops in the afterglow light curves  \S \ref{Plateaus};  (iii)  afterglow flares, \S \ref{flares1}.  For numerical estimates,  we assume the redshift $z = 1$, the Lorentz factor of the wind $\gamma_w = 5 \times 10^5$, the wind luminosity $L_w = 10^{46}$ erg/s, the initial injection time $t_0^{\prime} = 10^5$s (in jet frame), the power law index of particle distribution $p=2.2$, $\Gamma_{CD} \approx \gamma_{RS}$, and the viewing angle is 0 (observer on the axis)  for all calculations.  

\subsection{Plateaus and  sudden intensity drops in afterglow light curves}
\label{Plateaus}

\begin{figure}[h!]
  \centering
  \includegraphics[width=0.99\textwidth]{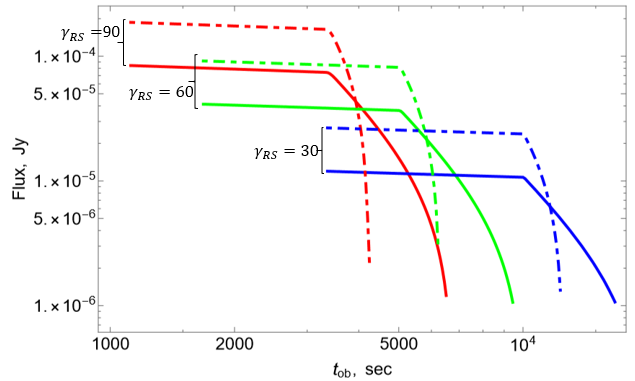}
  \caption{The light curve at 100 KeV  for different  Lorentz factors of the post-RS flow and different  jet angles $1/\gamma_{RS}$  (solid lines) and $1/2\gamma_{RS}$  (dotted-dash lines). Note that for $\theta_ j  < 1/\gamma_{RS} $  the drop in intensity is extremely fast. 
  }
  \label{GammaCD}
 \end{figure}

 Particles accelerated at the RS emit   in the fast cooling regime. The resulting synchrotron luminosity $L_s$ is approximately proportional to the wind luminosity $L_w$, as discussed by \cite{2017ApJ...835..206L}. (For highly magnetized winds with $\sigma\gg 1$ the RS emissivity is only mildly suppressed, by  high magnetization, $\propto 1/\sqrt{\sigma}$, due to the fact that higher sigma shocks propagate faster with respect to the wind.) Thus, the constant wind will produce a nearly constant light curve:  plateaus are natural consequences in our model in the case of constant long-lasting wind, see Fig. \ref{GammaCD}. 
   At the early times  all light curves show a nearly constant evolution with time, a plateau,  with flux $\propto t_{ob}^{-0.1}$. A slight temporal decrease is due to the fact that \Bf\ at the RS decreases with time so that particles emit less efficiently.  This observed temporal decrease is flatter than  what is typically observed, $\propto t_{ob}^{-\alpha_2}$ with $\alpha_2=0.5-1$  \citep{2006ApJ...642..389N}. A steeper decrease can be easily accommodated due to  the decreasing wind power. This explains the plateaus.


Next we assume that  the central engine suddenly  stops operating. This process could be due to the collapse of a \NS\ into a black hole or sudden depletion of an accretion disk. At a later time, when the ``tail'' of the wind reaches the termination shock, acceleration stops. Let the 
 injection terminate  at a some  time $t_{\text{stop}}^{\prime}$. The distribution function in the shocked part of the wind then become
\begin{eqnarray}
{N}(\gamma^{\prime},t^{\prime}) \propto \int _{t_0^{\prime}}^{\min(t^{\prime},t_{\text{stop}}^{\prime})}G(\gamma^{\prime},t^{\prime},t_i^{\prime})dt_i^{\prime}
\end{eqnarray}

\begin{figure}[h!]
  \centering
  \includegraphics[width=0.99\textwidth]{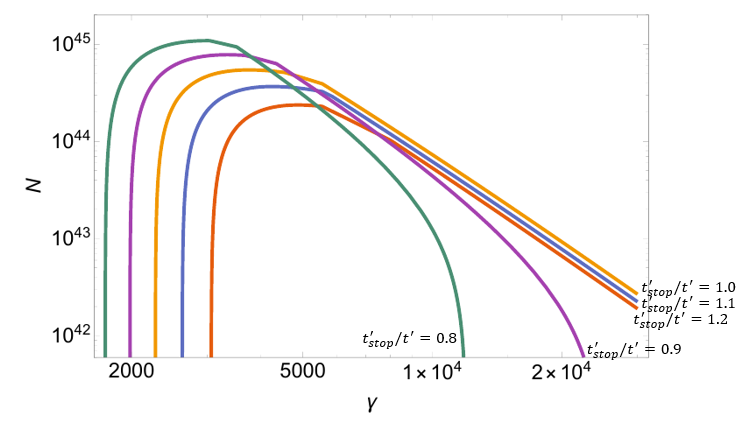}
  \caption{Evolution of the distribution function. Here we take  account the effect of radiation loss and adiabatic expansion. In our calculation, the Lorentz factor of RS $\gamma_{RS}=90$, and the injection is stopped at time $t_{\text{stop}}^{\prime} = 1.5 \times 10^5$s, $\gamma_{\min} = \gamma_w/\gamma_{RS} = 5556$, initial magnetic field $B_0 = 2.1$G. The times are measured in fluid frame at $t_{\text{stop}}^{\prime}/t^{\prime} = 1.2, 1.1, 1.0, 0.9, 0.8$ from red to green curves.}
  \label{stop_at_150000}
 \end{figure}

 Fig. \ref{stop_at_150000} shows the evolution of the distribution function by  assuming the Lorentz factor of RS $\gamma_{RS}=90$, and the injection is stopped at time $t_{\text{stop}}^{\prime} = 1.5 \times 10^5$s (in this case, the $T_{\text{ob,stop}} = 833$s in the observer's frame).  The number of high energy particles drops sharply right after  the injection is stopped:   particles lose their energy via synchrotron radiation and adiabatic expansion in fast cooling regime.

The resulting light curves   are plotted in Fig. \ref{GammaCD}. We assume post-RS flow $\gamma_{RS}=30, \, 60, \, 90$ and two jet opening angles of $\sim  \gamma_{RS}^{-1}$ and  $\sim (1/2)  \gamma_{RS}^{-1}$. These particular  choices of $\theta_j$ are  motivated by our expectation that sudden switch-off of the acceleration at the RS will lead to fast decays in the observed flux (in the fast cooling regime).

 The injection is stopped at a fixed time in the fluid frame,  corresponding to $t_0'=6 \times 10^5$s. 
   There is a sudden drop of intensity when the injection is stopped   ($T_{\text{ob}} = 10000$s for blue curve, $T_{\text{ob}} = 5000$s for green curve, and $T_{\text{ob}} = 3333$s for red curve).  Blue curve has $\gamma_{RS} = 30$, $\gamma_{\min} = \gamma_{\text{w}}/\gamma_{RS}= 16667$, initial magnetic field $B_0$ = 6.4G; green curve has $\gamma_{RS} = 60$, $\gamma_{\min} = \gamma_w/\gamma_{RS} =8333$, initial magnetic field $B_0$=3.2G; red curve has $\gamma_{RS} = 90$, $\gamma_{\min} = \gamma_w/\gamma_{RS}= 5556$, initial magnetic field $B_0$ = 2.1G. Here we assume $B_0 \propto {\gamma_{RS}}^{-1}$ for our calculations. Smaller jet angle produce sharper drop.  

In the simplest qualitative explanation, consider a shell  of radius $r_{em}$ extending to a finite angle $\theta_j$ and producing an instantaneous flash of emission (instantaneous is an approximation to the fast cooling regime). The observed light curve is then \cite{1996ApJ...473..998F}
\be 
\propto 
\left\{
\begin{array}{cc}
\left( \frac{T_{ob}}{T_0} \right)  ^ {-(\alpha +2)}, & 0< T_{ob} <  \frac{r_{em} /c} {2 } \theta_j^2 \\
0 &   \frac{r_{em} /c} {2 } \theta_j^2  <  T_{ob}
\end {array}
\right.
\ee
where $T_0 =  \frac{r_{em} /c} {2 \gamma_{RS}^2}$ and $\alpha$ is the spectral index. Thus,  for $\theta_j> 1/\gamma_{RS}$ the observed duration of a pulse is $\sim T_0$, while  for $\theta_j<  1/\gamma_{RS}$ the pulse lasts  much shorter, $\sim T_0 ( \theta_j \gamma_{RS})^2 \ll T_0$. Thus,  in this case 
a drop in intensity is faster than what would be expected in either faster  shocks or shocks producing emission in slowly cooling regime.

\subsection{Afterglow  flares}
\label{flares1}

\begin{figure}[h!]
  \centering
  \includegraphics[width=0.99\textwidth]{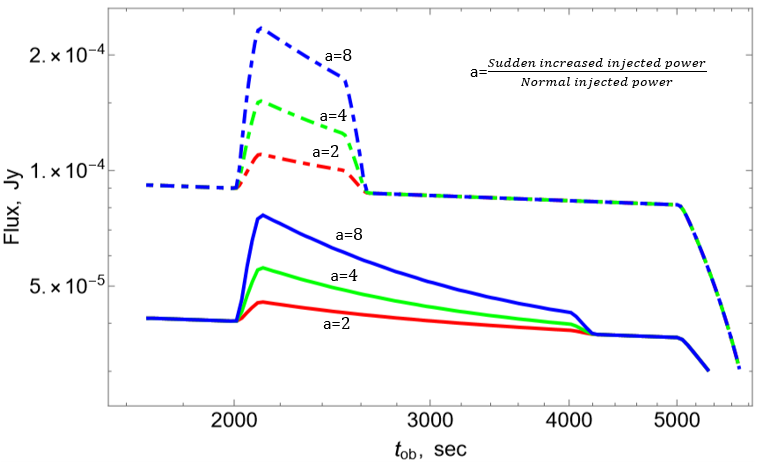}
  \caption{Afterglow flares due to variations in wind luminosity for the case  $\gamma_{RS} =60$ (green curve in the Fig. \protect\ref{GammaCD}). The ejected power is increased by factors  $a=2, 4, 8$ for a short period from $2.4 \times 10^5$s to $2.5 \times 10^5$s (in the fluid frame). Solid lines are for  $\theta_j = \gamma_{RS}$, dashed lines are for  $\theta_j = 1/2\gamma_{RS}$.
The relative shift of intensities between the plots for two opening angles is due to our parametrization of the injected power (constant total power).
}
  \label{flare}
 \end{figure}

Next, we investigate the possibility that  afterglow  flares are produced due to the variations in wind power. 
We re-consider the case of $\gamma_{RS} = 60$ (the green curve in Fig. \ref{GammaCD}),
  but set the ejected power at two, four, and eight times larger than the  average   power for a short period of time from $2.4 \times 10^5$s to $2.5 \times 10^5$s. We consider the two cases: the wide jet angle ($\theta_j = 1/\gamma_{RS}$) and the narrow jet angle ($\theta_j = 1/2\gamma_{RS}$). The corresponding light curves are plotted in Fig. \ref{flare}.

Light curves show a  sharp rise around $T_{ob} = 2000$ corresponding  to the increased ejected power $t = 2.4 \times 10^5$s at emission angle $\theta = 0$, followed by a   sharp drop around $T_{ob} = 4000$s for the case of wide jet and $T_{ob} = 2500$s for the case of narrow jet (which corresponds to the ending time of the increased ejected power $t = 2.5 \times 10^5$s at emission angle $\theta = \theta_j$). Bright flares are clearly seen. Importantly,   the corresponding total injected energy  is only  $\sim 1\%, \, 5\%$ and  $10\%$  larger than the averaged value.
 The magnitude of the rise in flux is less than the magnitude of the  rise in ejected power (e.g. the rise in ejected power by a factor eight only gives the rise in flux by a factor two), due to the fact that   the emission from the increased ejected power from different angles is spread out in observer time.
Thus, variations in the wind power, with  minor total energy input, can produce bright afterglow flares.

\section{Discussion}
\label{discussion}

In this paper, we calculate emission properties expected from (non-stationary) particle injection at the termination shock of long-lasting GRB engines. We assume  a ``pulsar paradigm'': the central engine produces  ultra-relativistic, highly magnetized wind with particles  accelerated at the wind termination shock \citep{kennel_coroniti_84b}. 
\citep[In contrast,  a number of authors, \eg,][discussed  long-lasting engine that produces colliding  shells, in analogy with the fireball model for the  prompt emission]{1994ApJ...430L..93R}. 

The key advantages of the model are high radiation efficiency, and the ability to produce fast temporal variations.
We can reproduce
\begin{itemize} 
\item  Afterglow plateaus: in the  fast cooling regime the emitted power is comparable to the wind power. Hence, only mild wind luminosity $L_w\sim 10^{46}$ erg s$^{-1}$ is required
\item Sudden drops in afterglow curves: if the central engine stops operating, and if at the corresponding moment the \Lf\ of the RS is of the order of the jet angle, a sudden drop in intensity will be observed.
\item Afterglow flares: if the wind intensity varies, this leads to the sharp variations of afterglow luminosities. Importantly, a total injected {\it energy} is small compared to the total energy of the explosion.
\end {itemize} 

Our model may provide explanations to  other problems in GRBs' afterglow. (i) ``Naked GRBs problem'' \citep[][]{2006ApJ...637L..13P,2008AIPC.1000..191V}:  if the explosion does not produce a long-lasting wind, then there will be no X-ray afterglow since RS reflects the properties of wind. (ii)   ``Missing orphan afterglows'':
both prompt emission and afterglow emission arise from the engine-powered flow, so they may  have similar collimation properties.

Finally, let us comment on  another conceptual point: analytical and numerical studies of relativistic double explosions  \citep{2017ApJ...835..206L,2017PhFl...29d7101L,2020arXiv200413600B} assumed that the initial FS has reached a self-similar \cite{BlandfordMcKee} stage. This is an important (and not fully justified)  simplification: in reality one expects that the dynamics of the second set of shocks will be influenced 
by the density structure of ejecta, resulting in  shell-induced variations of the \Lf\ of the contact discontinuity and of the reverse shock. This will produce additional  variations of the  RS emissivity.

This research was supported by NASA Swift grant 1619001.

\bibliographystyle{apj}
 \bibliography{/Users/maxim/Home/Research/BibTex,reference}

\end{document}